\documentclass[11pt,a4paper]{article}
\usepackage{graphicx}
\usepackage{amssymb}
\usepackage{amsmath}
\usepackage{./cite}

\newcommand{\bm}[1]{\mbox{\boldmath$#1$}}

\title{Distribution of the distance between opposite nodes of random
polygons with a fixed knot}
\author{$^1$Akihisa Yao, $^2$Hiroshi Tsukahara, $^3$Tetsuo Deguchi and $^1$Takeo Inami\\
\\
$^1$Department of Physics, Faculty of Science and Engineering,\\
Chuo University,
1-13-27 Kasuga, Bunkyo-ku, Tokyo 112-8551\\
\\
$^2$Geographic Information Systems Department,\\
Hitachi Software Engineering Co., Ltd.\\
4-12-7 Higashishinagawa, Shinagawa-ku, Tokyo 140-0002\\
\\
$^3$Department of Physics,\\
Faculty of Science, Ochanomizu University\\
2-1-1 Ohtsuka, Bunkyo-ku, Tokyo 112-8610}
\date{June 7, 2004}

\begin{document}
{
\renewcommand{\arraystretch}{0.5}
\maketitle
}

\begin{abstract}
	%Abstract
%Abst.tex
%2004/02/16 Created

We examine numerically the distribution function $f_K(r)$ of 
distance $r$ between opposite polygonal nodes 
for random polygons of $N$ nodes with a fixed knot type $K$. Here we consider  
three knots such as $\emptyset$, $3_1$ and $3_1 \sharp 3_1$. 
 In a wide range of $r$, the shape of $f_K(r)$ 
is well fitted by the scaling form~\cite{DES_CLOIZEAUX.1974.01} 
%derived from the renormalization group theory 
of self-avoiding walks. 
The fit yields the Gaussian exponents $\nu_K = {1 \over 2}$ and $\gamma_K = 1$. 
Furthermore, if we re-scale the intersegment distance $r$ by the 
average size $R_K$ of random polygons of knot $K$,   
the distribution function of the variable $r/R_K$ should become 
the same Gaussian distribution for any large value of $N$ and any knot $K$. 
We also introduce a fitting formula 
to the distribution $g_K(R)$ of gyration radius $R$ 
for random polygons under some topological constraint $K$.

\end{abstract}

\section{Introduction}
%Introduction
%Intro.tex
%2004/02/16 Created
%2004/02/24 Last Modified Date

Polymer chains in solutions or gels 
may be highly self-entangled:
such entanglements should be important to understand
some features of polymeric materials.
A variety of knots can appear by connecting the two ends of a polymer chain.
In fact, various knotted DNAs are synthesized in experiments 
through random closure of nicked DNA chains 
~\cite{SHAW-WANG.1993.01,RYBENKOV-COZZARELLI-VOLOGODSKII.1993.01}.  
Since topological questions 
were addressed by Delbr\"uck, Frisch and Wasserman~\cite{DELBRUCK.1962.01,FRISCH-WASSERMAN.1961.01},
several aspects of knotted ring polymers,  
such as the probability of random
knotting~\cite{VOLOGODSKII-ET-AL.1974.01,MICHELS-WIEGEL.1982.01,DES_CLOIZEAUX-MEHTA.1979.01,SUMNERS-WHITTINGTON.1988.01,PIPPENGER.1989.01,KONIARIS-MUTHUKUMAR.1991.01,DEGUCHI-TSURUSAKI.1994.01,DEGUCHI-TSURUSAKI.1997.02,SHIMAMURA-DEGUCHI.2001.01,DOBAY-SOTTAS-DUBOCHET-STASIAK.2001.01}, the average sizes ~\cite{ORANDINI-TESI-JANSE_VAN_RENSBURG-WHITTINGTON.1998.01,SHIMAMURA-DEGUCHI.2002.01}
and the complexity of their conformations~\cite{SHIMAMURA-DEGUCHI.2003.01}
have been studied numerically and analytically.

Let us discuss the average size of knotted ring polymers with no
excluded volume, i.e.,  
the mean-squared gyration radius $R^2_K(N)$ of $N$-noded random polygons   
with fixed knot type $K$    
~\cite{DES_CLOIZEAUX.1981.01,DEUTSCH.1999.01,GROSBERG.2000.01,SHIMAMURA-DEGUCHI.2001.03,MATSUDA-YAO-TSUKAHARA-DEGUCHI-FURUTA-INAMI.2003.01,DOBAY-DUBOCHET-MILLET-SOTTAS-STASIAK.2003.01,GROSBERG.2003.01}. 
We consider random polygons as a simple model of ring polymers 
in solution at the $\theta$-point~\cite{DE_GENNES.1979.01}. 
At the $\theta$-point polymers should have no effect of excluded volume. 
Furthermore, ring polymers keep their topology unchanged. 
It has now been established in simulations 
~\cite{DEUTSCH.1999.01,SHIMAMURA-DEGUCHI.2001.03,MATSUDA-YAO-TSUKAHARA-DEGUCHI-FURUTA-INAMI.2003.01,DOBAY-DUBOCHET-MILLET-SOTTAS-STASIAK.2003.01}  
that  the average size of random polygons with a fixed knot
is larger than that with no topological constraint, 
when $N$ is large. 
The topological swelling of random polygons may be explained in terms of  
entropic repulsion caused by the topological constraint.  
The phenomenon should be closely related to the ``topological excluded volume" 
proposed for such random polygons that possess  the trivial knot $\emptyset$
~\cite{DES_CLOIZEAUX.1981.01}.
Concerning the large-$N$ behavior of $R^2_K(N)$,   
however, the numerical studies do not unanimously arrive 
at the same conclusion. 
We have analyzed the data of $R^2_K(N)$ for a model of random polygons
~\cite{MATSUDA-YAO-TSUKAHARA-DEGUCHI-FURUTA-INAMI.2003.01}, 
assuming the scaling formula of the following form:  
\begin{equation}
R_K^2(N) = A_K N^{2 \nu_K} \left( 1 + B_K N^{-\Delta_K} + \ldots \right) \, . 
\label{eq:size_scaling}
\end{equation}
The result favors to the interpretation of   
$\nu_K = \nu_{\rm SAW}$. However, limiting the analysis to a narrower 
range of $N$, the alternative interpretation $\nu_K = \nu_{\rm  RW}$ 
is also possible.  
Here, self-avoiding walks (SAW) and random walks (RW) 
have the scaling exponent 
$\nu_{\rm SAW} = 0.588$ and $\nu_{\rm RW}=0.5$, respectively.  
%
%It seems that it is nontrivial to determine 
%the asymptotic exponent $\nu_K$ numerically. 
%
%Furthermore, even other possibilities have not been completely 
%excluded, yet. 
Thus, in order to clarify the large-$N$ behavior 
of random polygons with fixed knots,    
it would  be interesting to investigate some other quantity 
associated with the asymptotic behavior.

In this paper we study the following two quantities 
of random polygons with a fixed knot:  
i) the distribution function of the distance between opposite nodes 
and ii) the distribution of the radius of gyration.  
If the ``topological excluded volume" corresponds   
to a certain amount of excluded volume, 
then  the distance between opposite nodes 
should follow a non-Gaussian distribution.      
 For ring polymers with excluded volume, 
the distribution of the distance between opposite nodes 
should be non-Gaussian, 
while it is Gaussian for random polygons. 
Here we assume that the distance between opposite polygonal nodes 
plays the similar role as the end-to-end distance of a linear chain.  
 For the self-avoiding walk, the end-to-end distance distribution 
is non-Gaussian~\cite{BISHOP-CLARKE.1991.01,BISHOP-CLARKE.1991.02}. 
%Its scaling behaviors have been discussed in terms of field theories. 

Through computer simulation of random polygons with fixed knots, 
we have evaluated the distributions of the distance $r$ 
between opposite nodes of random polygons under the  
topological constraints \cite{YAO.2004.01}.  
We are concerned with the trivial knot, the trefoil knot, 
and the composite knot consisting of two 
trefoil knots, which are denoted by  
$\emptyset$, $3_1$, and $3_1 \sharp 3_1$, respectively. 
We show that the scaling form of the self-avoiding walk gives good fitting curves to the
data of the distributions in a wide range. 
It should be remarkable, since the scaling form 
of the end-to-end distance distribution is derived  when 
$\rho= r / N^{\nu}$ is finite and very large 
~\cite{FISHER.1966.01,MCKENZIE-MOORE.1971.01,DES_CLOIZEAUX.1974.01,LIPKIN-OONO-FREED.1981.01,
BISHOP-CLARKE.1991.01,BISHOP-CLARKE.1991.02,CARACCIOLO-CAUSO-PELISSETTO.2000.01,LUE-KISELEV.1999.01}. 
Furthermore, we show that the distribution function of normalized distance $r/R_K$ 
should be given by the same Gaussian form for any $N$ and $K$. 
Here we recall that $R_K$ denotes the average size of random polygons of knot $K$.  
Thus, it is suggested that the effect of the 
``topological excluded volume'' should be 
 different from the standard excluded-volume effect. 

%. In fact, the scaling form of the end-to-end distance distribution of
% SAW is derived in a limited range by taking the limit where the length $N$ 
% and the distance $r$ goes to infinity keeping the normalized distance 
%$\rho= r / N^{\nu}$ finite and very large 

We have also evaluated the distribution of the gyration radius 
for random polygons under some topological constraints~\cite{YAO.2004.01}.  
We introduce a formula for describing the distribution, 
and discuss its fitting curves.  
The formula of the gyration-radius distribution 
is new, in particular, for random polygons 
under topological constraints.  
We note that  
 for the Gaussian random walk, several approximate formulas of the gyration-radius 
distribution are known~\cite{FIXMAN.1962.01,YAMAKAWA.1971.01}. 

%
% Compared to the end-to-end distribution function, however, 
%scaling properties of the gyration-radius distribution have not been 
%extensively discussed, yet, either theoretically or even numerically.  
%For instance, it is not known whether there exist 
%connections between a certain asymptotic behavior of the gyration-radius
% distribution and the critical exponent of a macroscopic quantity. 
%

The paper consists of the following.  In \S 2, we
explain the model of random polygons, and define some symbols 
for the distribution functions.  In \S 3, we describe briefly 
some procedures of the computer simulation. 
In \S 4, we discuss the numerical results of the present research.  
 We plot the distribution function of the intersegment distance 
for random polygons under topological constraints.  
Through fitting curves to the data, we discuss that the distribution 
of the intersegment distance should be well approximated 
by the Gaussian distribution. We also plot the distribution of the gyration radius 
for random polygons under topological constraints. 
We thus investigate topological effects on the average sizes of random polygons.   
In \S 5, we discuss that the ``topological excluded volume'' should be different 
from the standard excluded volume, and present an open question.

\section{Model and distribution functions}
%Model
%model.tex
%2004/02/16 Created
%2004/02/18 Last Modified Date

%\subsection{Model}

We consider a model of random polygons 
in which a polygon ${\mathcal P}_N$ consists of $N$ line
segments of length $a$.  It is specified  
by position vectors of its nodes, 
${\mathcal P}_N = \left( {\bm r}_1, {\bm r}_2, \dots, {\bm r}_N \right)$. 
All cyclic permutations of the set of position vectors 
correspond to the same polygon. 
We recall that random polygons have no excluded volume.      
Hereafter we set $a=1$. 
%The vectors of ${\mathcal P}_N$ satisfy the geometrical constraint 
%$|r_{i+1}-r_{i}| = a$ for $1 \leq i \leq N$ where we use the convention $r_{N + i} = r_i$.
%they consist of purely geometric points and edges. 
% to unity since the scaling behavior is independent 
%of the macroscopic details of the model. 

When a polygon is topologically equivalent to a knot $K$, 
we call it a polygon of knot type $K$.
The configuration space $\mathcal{C}$ of polygons is divided into subspaces 
$\mathcal{C}_K$ in which all polygons have the same knot $K$.  
We have $\mathcal{C} = \sum_K \mathcal{C}_K$.

%\subsection{Distribution of the distance between two nodes }

For a polygon ${\mathcal P}_N$, we denote the intersegment vector 
from the $i$th node to the $i+ {\lambda}N$th node 
%from ${\bm r}_i$ to ${\bm r}_{i+{\lambda}N}$ as
\begin{equation}
{\bm r}(i;\lambda,{\mathcal P}_N) 
= {\bm r}_{i+{\lambda}N} - {\bm r}_i,
\label{eq:intersegment_vector}
\end{equation} 
where the progress parameter $\lambda$ takes a value between 0 and 1.
Here we assume the convention: ${\bm r}_{N + i} = {\bm r}_i$.

We define the distribution 
of the distance between the $i$th and the $i+ {\lambda}N$th nodes 
by the probability  $f(r;\lambda,N) \Delta r$ that 
the length of the intersegment vector 
${\bm r}(i;\lambda,{\mathcal P}_N)$ takes a value between
$r$ and $r+\Delta r$: 
\begin{equation}
f(r;\lambda,N) \Delta r = \frac{1}{NM} \sum_{m = 1}^{M} 
\sum_{i = 1}^{N} \int_{r}^{r+\Delta r} dr\, 
\delta \left( r - |{\bm r}(i;\lambda,{\mathcal P}_{N,m})| \right).
\label{eq:dist_inter_all}
\end{equation}
Here $\Delta r$ is a small positive real number.  
We choose it so that the statistical fluctuation of $f(r;\lambda,N)$ 
becomes moderately small. 
The distribution of the distance between two nodes 
for random polygons with a fixed knot type $K$ is similarly defined  by 
\begin{equation}
f_K(r;\lambda,N) \Delta r = \frac{1}{NM_K} \sum_{m = 1}^{M} 
\sum_{i = 1}^{N} \int_{r}^{r+\Delta r} dr\, 
\delta\left( r - |{\bm r}(i;\lambda,{\mathcal P}_{N,m})| \right)
\chi({\mathcal P}_{N,m},K),
\label{eq:dist_inter_K}
\end{equation}
Here the indicator function $\chi({\mathcal P},K)$ 
filters the polygons of knot type $K$;
it takes the value $1$ if ${\mathcal P} \in \mathcal{C}_K$ and 0 otherwise.

We calculate the distribution $f(r;\lambda,N)$ 
of the intersegment distance $r$ by randomly generating 
a large number of polygons ${\mathcal P}_{N,m}$ with length $N$ 
for $m=1, \ldots, M$.  
Here the subscript $m$ denotes the $m$th polygon generated.
The number of generated polygons of the knot type $K$ is given by  
$M_K = \sum_m \chi({\mathcal P}_{N,m}, K)$,  
and we have $M = \sum_K M_K$.

%\subsection{Distribution of the gyration radius}

Let us denote the square of the gyration radius 
of a polygon ${\mathcal P}_N$  by 
\begin{equation}
R_G^2({\mathcal P}_N) = \frac{1}{2N^2} \sum_{i,j=1}^{N} 
\left( {\bf r}_i - {\bf r}_j \right)^2.
\label{eq:radius_of_gyration}
\end{equation}
We define the distribution $g(R;N)$ for gyration radius $R$ by   
\begin{equation}
g( R; N ) \Delta R = \frac{1}{M} \sum_{m = 1}^{M} \int_{R}^{R+\Delta R} dR\, 
\delta\left( R - \sqrt{R_G^2({\mathcal P}_{N,m})} \right),
\label{eq:dist_size_all}
\end{equation}
and the one for polygons with knot type $K$ by
\begin{equation}
g_K( R; N ) \Delta R = \frac{1}{M_K} \sum_{m = 1}^{M} 
\int_{R}^{R+\Delta R} dR\, 
\delta\left( R - \sqrt{R_G^2({\mathcal P}_{N,m})} \right)
\chi({\mathcal P}_{N,m},K).
\label{eq:dist_size_K}
\end{equation}

\section{Simulation procedure}
%Simulation procedure
%Sim.tex
%2004/02/16 Created
%2004/02/18 Last Modified Date

A pivot move for a polygon is a rotation of a chain of segments, randomly chosen from the polygon,
around the axis passing the two endmost nodes of the chain by a random
amount of angle $\phi$ 
\cite{FREIRE-HORTA.1976.01,MATSUDA-YAO-TSUKAHARA-DEGUCHI-FURUTA-INAMI.2003.01}.
The rotation angle $\phi$ is selected randomly from the interval between $0$ and $360$ degrees.
We do not check self-intersections during the process of rotation of the chain 
since such configurations are negligible in the space ${\mathcal C}$.

We start from a seed conformation placed on the cubic lattice, which is regarded as
a special conformation of the off-lattice polygon 
in the continuum space~\cite{YAO-MATSUDA-TSUKAHARA-SHIMAMURA-DEGUCHI.2001.01}.
We then generate a sequence of polygons by applying the pivot moves repeatedly.
After discarding the initial $2000$ transient conformations,
we take samples of polygons at every $200$ pivot moves.

To determine the topology of polygons, we employ two simple knot invariants.
We calculate the special value of the Alexander polynomial
$\Delta_K(t)$ at $t = -1$~\cite{VOLOGODSKII-ET-AL.1974.01} 
(which is also called the determinant of a knot),
and the Vassiliev invariant of the second order $v_2(K)$
~\cite{DEGUCHI-TSURUSAKI.1993.01,POLYAK-VIRO.1994.01}.
With these invariants, the chance of misidentification of 
topology class for a given  polygon 
should be negligible and much  smaller than the statistical errors 
of the data, as far as the simple knots are concerned.

The simulation has been performed for polygons with the length $N = 300$ 
and $600$.
We have generated $M = 3 \times 10^6$ random polygons 
for each given length $N$.
We have classified those polygons into four groups according 
to their knot types, the three groups of polygons with the specific knot types 
%such as the trivial knot 
$\emptyset$, 
%the trefoil knot 
$3_1$, and 
% the composite knot 
$3_1\sharp3_1$, 
and the other  group of knot types other than the previous three. 
The three knots have distinct sets of values for the two knot invariants 
$|\Delta_K(t=-1)|$ and $v_2(K)$.

The distribution function $f_K( r; \lambda, N)$ of the intersegment
distance $r$ has been evaluated 
at the progress parameter $\lambda = 1/4$, $1/2$ and $3/4$ for 
random polygons under some topological constraint $K$ \cite{YAO.2004.01}. 
However, we focus on the case of $\lambda=1/2$. 
%The other cases will be discussed in a later publication. 
The range of intersegment distance $r$ 
is divided into a number of bins of with the width $\Delta r$. Here we
set $\Delta r = 0.25$. We enumerate the number
of  intersegment distances in each of the bins.
The distribution function is  obtained by dividing the number of
each bin by the total number of intersegment distances.
Similarly we numerically evaluate the distribution $g_K(R;N)$ of
gyration radius $R$ for random polygons under some topological constraint $K$. 
Here we take $\Delta R = 0.25$.

\section{Results of the simulation}
%Results
%Results.tex
%2004/02/16 Created
%2004/02/18 Last Modified Date

%Approach to analysis
%Approach.tex
%2004/02/16 Created
%2004/02/18 Last Modified Date

%We are concerned with the asymptotic behaviors of 
%the distribution $f_K(r;\lambda,N)$ in the scaling limit 
%where $N$ and $r$ grows infinitely 
%but the ratio $\rho = r/R_K(N)$ is kept finite.

\subsection{Functional forms of the distributions }

The asymptotic scaling form of the end-to-end distance 
distribution of the self-avoiding walks 
is derived for the region $\rho=r/N^{\nu} \gg 1$ 
~\cite{FISHER.1966.01,MCKENZIE-MOORE.1971.01,DES_CLOIZEAUX.1974.01}.
We now apply it to the data of the distribution function 
for the distance between opposites nodes for random polygons 
under topological constraint $K$. We thus 
have the following: 
\begin{eqnarray}
	f_K(r;\lambda,N) & = & A_K r^{2 + \theta_K} 
	\exp\left[-D_K r^{\delta_K}\right],
	\label{eq:scaling_large_dist_inter_K}
	\\
	\theta_K &=& {d \nu_K + 1 - \gamma_K - d/2 \over 1 - \nu_K},
	\label{eq:theta_large_dist_inter_K}
	\\
	\delta_K &=& {1 \over 1 - \nu_K} .
	\label{eq:delta_large_dist_inter_K}
\end{eqnarray}
Hereafter we set $d=3$. 

 For the distribution $g_K(R;N)$ of gyration radius 
$R$, we introduce the following formula: 
\begin{equation}
	g_K(R;N) = A_{g,K} |R - c_K|^{\theta_{g,K}}
		\exp \left[ -D_{g,K} |R - c_K|^{\delta_{g,K}} \right] \, . 
\label{eq:scaling_dist_size_K}
\end{equation}
For the Gaussian random walk there are some approximate expressions for  
the distribution of the gyration radius ~\cite{FIXMAN.1962.01,Flory-Fisk}. 
(See also \S 8 of Ref. \cite{YAMAKAWA.1971.01}.) 
For instance, the large $R$ case of Fixman's result~\cite{FIXMAN.1962.01} corresponds 
to a special case of the formula (\ref{eq:scaling_dist_size_K}), where we set 
$\delta_{g,K}=2$, $c_K=0$, and $\theta_{g,K}= 1$.

\subsection{Distribution function $f_K(r; \lambda, N)$
of intersegment distance $r$ }

\begin{figure}[htbp]
    	\includegraphics[width=130mm]
{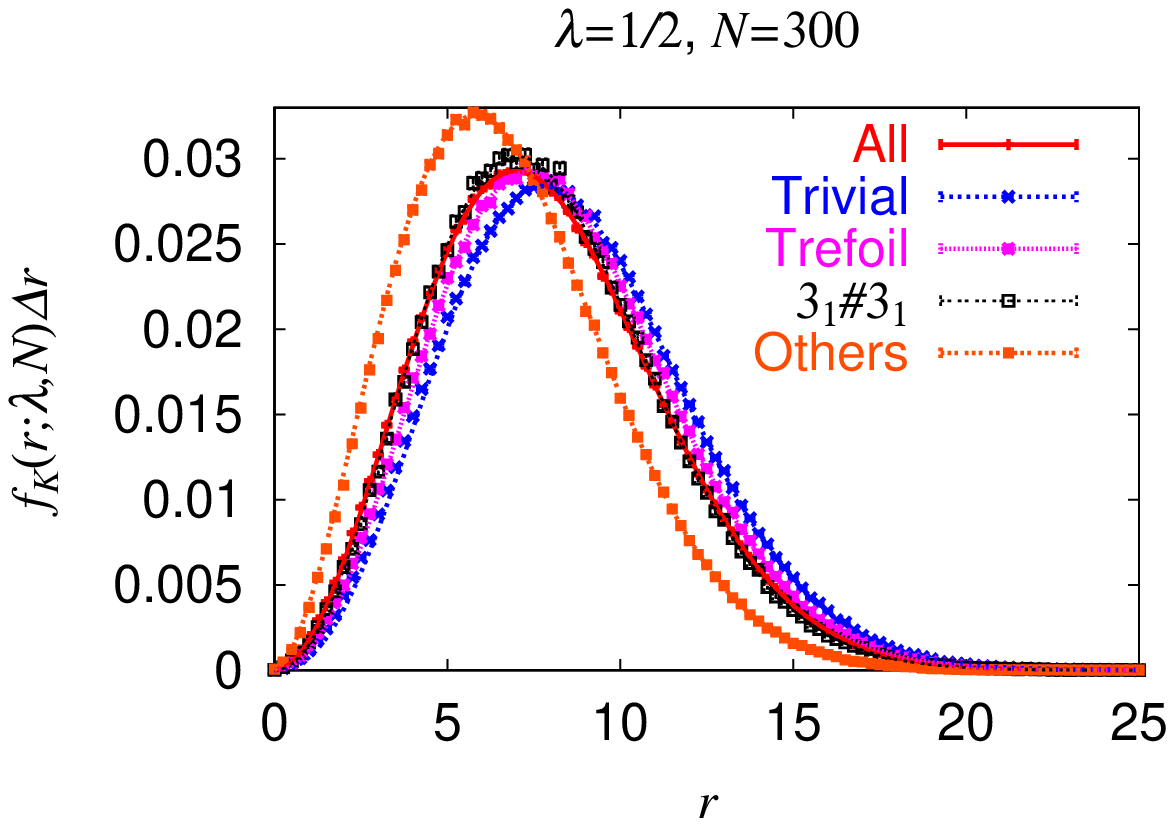}
    	\caption{Distribution $f_K(r;\lambda,N)$ 
    	of intersegment distance $r$ at $\lambda = 1/2$ for $N = 300$.
        Dots($\centerdot$), crosses($\times$), double crosses($\ast$), 
        open squares($\square$) and closed squares($\blacksquare$)
        denote the plots of conditions, $all$, $\emptyset$, $3_1$,
        $3_1\sharp 3_1$, and $others$, respectively.
        The plots and fitting curves for $all$, $\emptyset$, $3_1$,
        $3_1\sharp 3_1$, and $others$ are colored with
        red, blue, fuchsia, black and orange, respectively.
        }
		\label{fig:dist_inter_half_300}
\end{figure}

\begin{figure}[htbp]
    	\includegraphics[width=130mm]
{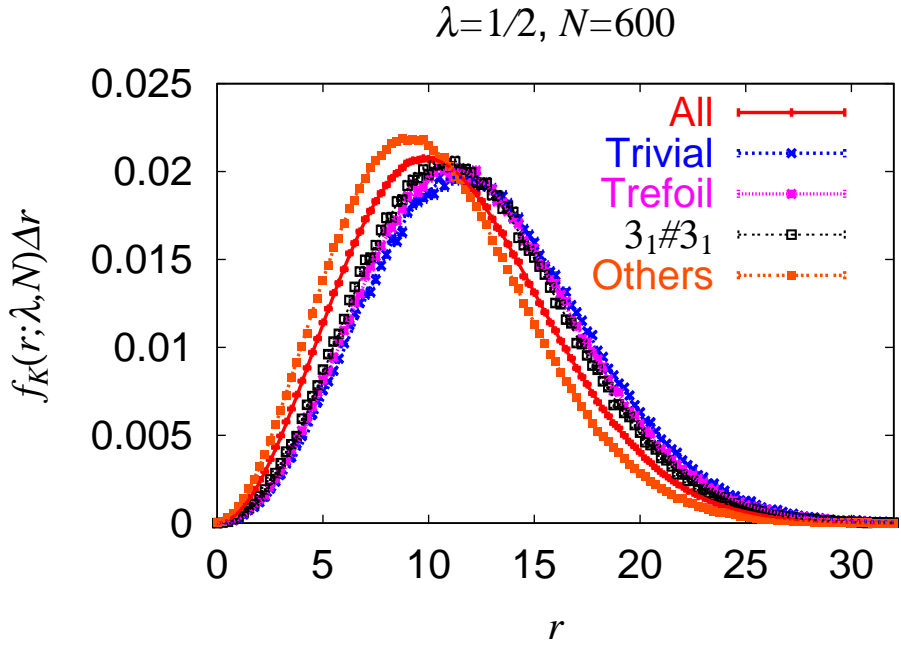}
    	\caption{Distributions $f_K(r;\lambda,N)$ 
    	of intersegment distance $r$ at $\lambda = 1/2$ for $N = 600$.
	Dots($\centerdot$), crosses($\times$), double crosses($\ast$), 
        open squares($\square$) and closed squares($\blacksquare$)
        denote the plots of conditions, $all$, $\emptyset$, $3_1$,
        $3_1\sharp 3_1$, and $others$, respectively.
        The plots and fitting curves are colored with
        red, blue, fuchsia, black and orange, respectively.
}
		\label{fig:dist_inter_half_600}
\end{figure}

The intersegment distributions $f_K(r; \lambda, N)$ at $\lambda=1/2$ 
for $N=300$ and $600$
are presented in figures~\ref{fig:dist_inter_half_300} and
\ref{fig:dist_inter_half_600}, respectively. 
Here the topological conditions denoted by $K$ 
correspond to restriction of random polygons  
into the following sets: 
(i) all polygons;  
(ii) polygons of the trivial knot $\emptyset$;  
(iii) polygons of the trefoil knot $3_1$;  
(iv) polygons of the composite knot $3_1 \sharp 3_1$; 
(v) polygons of any knot types other than the three knots 
$\emptyset$, $3_1$, and $3_1 \sharp 3_1$. 
We thus consider the five different topological conditions. 
We note that the case (i) corresponds to no topological constraint. 
We denote the distribution functions 
of the five cases simply as  $f_{all}$, $f_{\emptyset}$, 
$f_{3_1}$, $f_{3_1 \sharp 3_1}$, and $f_{others}$, respectively.

The fitting curves of figures~\ref{fig:dist_inter_half_300} 
and \ref{fig:dist_inter_half_600} are fit well to the data points.  
The curves are given by the scaling form 
(\ref{eq:scaling_large_dist_inter_K}), 
and are all very close to the Gaussian distributions.  
Here we note that it is also the case with the data for    
 $\lambda = 1/4$ and $3/4$. 
The numerical estimates for the exponents $\theta_K$ and $\delta_K$ 
and the constants $A_K$ and $D_K$ are given in 
table~\ref{table:fit_result_dist_inter}.  
The actual ranges of distance $r$ used for the fitting curves 
are also shown in table~\ref{table:fit_result_dist_inter}.   
The fitting curves fit very well to the data points not only 
in the range of $r$ larger than the peak position  
but almost in the entire range of $r$. 
The $\chi^2$ values per datum are very small.
Very small deviations are only seen in the small $r$ region, 
although the region is out of the fitting ranges.

The best estimates of the exponents $\theta_K$ and $\delta_K$ 
almost agree with the Gaussian values, i.e., 
$\nu_K \approx 1/2$ and $\gamma_K \approx 1$, 
within the range of estimation errors, 
for all the five different topological conditions 
and for both $N = 300$ and $600$.
 The constant $D_K$ depends on the polygonal length $N$. 
However, it does not change very much for the different knot types 
with respect to  the estimation errors. 
The constant $A_K$  depends on the knot type $K$ for $N = 300$.
However, the difference among $A_K$'s becomes smaller for $N = 600$ than for 
$N=300$. It is thus suggested that they should become 
the same value when $N$ is very large.

Let us denote by $r_K^{*}$ the peak position of the distribution $f_K(r)$. 
% The positions of peaks reflect their topological conditions. 
Assuming the scaling form (\ref{eq:scaling_large_dist_inter_K}),  
the peak position $r_K^{*}$ is given by   
% $f_K(r)$ have its largest value 
\begin{equation} 
r_K^{*} = \left( {\frac {2 + \theta_K } {D_K \delta_K}} \right)^{1 - \nu_K} 
\end{equation}
The peak position $r_K^{*}$ may characterize the knot dependence 
of the distribution function $f_K(r)$.    
When  $\nu_K=0.5$ and $\gamma_K=1.0$,    
the form of $f_K(r)$ is determined by the parameter $D_K$.  
%
%Here $A_K$ is derived from the normalization condition. 
%

In figure~\ref{fig:dist_inter_half_300}, the peak position $r_{\emptyset}^{*}$
of the distribution $f_{\emptyset}(r)$ is larger than  $r_{all}^{*}$ 
of $f_{all}$. 
For $f_{others}$,  $r_{others}^{*}$ is smaller than $r_{all}^{*}$. 
In figure~\ref{fig:dist_inter_half_600}, 
the peak positions of $f_{\emptyset}$, $f_{3_1}$, 
and $f_{3_1 \sharp 3_1}$ are all larger than that of $f_{all}$ for $N=600$. 
Their values of $N=600$ are 
much closer to each other than in the case of $N=300$.  
Here  the peak position $r_{others}^{*}$ of $f_{others}$ 
is smaller than that of $f_{all}$ also in the case of $N=600$. 
It is thus suggested that when $N$ is very large, 
the peak positions of $f_K(r)$ of simple knots  
should be given by the same value 
and the distributions $f_K(r)$ should approach a universal form.

The observations in figures~\ref{fig:dist_inter_half_300} and 
~\ref{fig:dist_inter_half_600} 
suggest that fixing a knot type of a random polygon 
leads to effective repulsion or attraction 
among internal segments of the polygon 
depending on the complexity of the knot type.
When the length $N$ becomes very large,   
polygons of very complex knots can appear.   
They should have smaller conformations 
than other polygons of simpler knots. 
As we see in figure~\ref{fig:dist_inter_half_300} for $N=300$, 
random polygons with the trivial knot  
have larger conformations on the average 
than those of no topological constraints,   
while random polygons of more complex knots 
have smaller conformations. 
This should be consistent with the effective swelling 
observed in the studies on the average sizes of 
random polygons with some fixed knots 
~\cite{DEUTSCH.1999.01,SHIMAMURA-DEGUCHI.2001.03,MATSUDA-YAO-TSUKAHARA-DEGUCHI-FURUTA-INAMI.2003.01,DOBAY-DUBOCHET-MILLET-SOTTAS-STASIAK.2003.01}.

%%%%%%%%%%%%%%%%%%%%%%%%%%%%%%%%%%%%
%	Normalized distance distribution
%	 for two points of polygons
%
%\vskip 24pt 
%

Let us discuss the $\lambda$- and $N$-dependence of the distribution 
function $f_K(r, \lambda, N)$ for a knot $K$.  
We denote by $r_K(\lambda,N)$ 
the average of the intersegment distance $r$  
at the parameter $\lambda$ for random polygons of $N$ nodes with the knot $K$. 
We shall suggest that for a given knot $K$, 
the distribution function ${f}_K(r, \lambda, N)$ 
should depend on $N$ and $\lambda$ only through the value 
$r_K(\lambda,N)$.

\begin{figure}[htbp]
    	\includegraphics[width=130mm]
{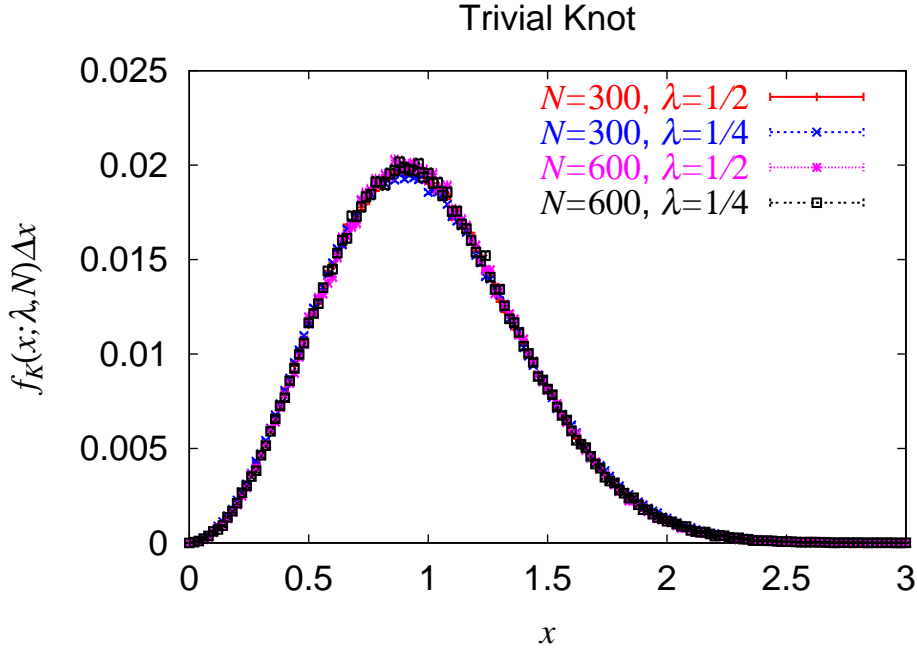}
    	\caption{Distributions ${\tilde f}_{\emptyset}(x; \lambda,N)$ 
    	of normalized intersegment distance 
    	$x=r/r_K$ at $\lambda = 1/2, 1/4$ and for 
    	$N = 300, 600$.
	Dots($\centerdot$), crosses($\times$), double crosses($\ast$)
        and open squares($\square$) denote the plots 
        with $(\lambda,N) = (1/2,300), (1/4,300), (1/2,600)$ and $(1/4,600)$, 
        respectively.
        They are displayed with red, blue, fuchsia and black, respectively.
Here $\Delta r$ of eq. (\ref{eq:dist_inter_K}) is given by $\Delta x = 0.02$.  
}
		\label{fig:universal}
\end{figure}

We introduce the distribution ${\tilde f}_K$ of 
normalized intersegment distance $x=r/r_K$. Here  we note   
${\tilde f}_K(x, \lambda, N) dx$ = $f_K(r, \lambda, N) dr $. 
The data for the three knots show that  
the function ${\tilde f}_K$ does not depend on either 
$\lambda$ or $N$. In figure~\ref{fig:universal} the data points of 
the distribution function ${\tilde f}_{\emptyset}(x; \lambda,N)$ 
of the normalized intersegment distance 
 $x=r/r_{\emptyset}$ are shown 
 for the four cases: $\lambda$ = 1/2 or 1/4 and  
 $N$ = 300 or 600. It is clear that 
the data points for all the four cases are located on the 
same curve.  
The best estimates of the fitting 
parameters to the data of ${\tilde f}_K$ are given 
in tables \ref{table:refit_result_1/2}
and \ref{table:refit_result_1/4} for $\lambda=1/2$ and 
1/4, respectively. As far as the five topological conditions are concerned, 
each of the fitting parameters of a condition $K$ has almost the same value for 
 $N$=300 or 600 and for $\lambda=1/2$ or 1/4. 
Thus, we suggest that for a given knot $K$ the distribution 
of the normalized intersegment distance, ${\tilde f}_K(x, \lambda, N)$, 
should be given by the same Gaussian form for any $N$ and $\lambda$.

\begin{figure}[htbp]
\includegraphics[width=130mm]{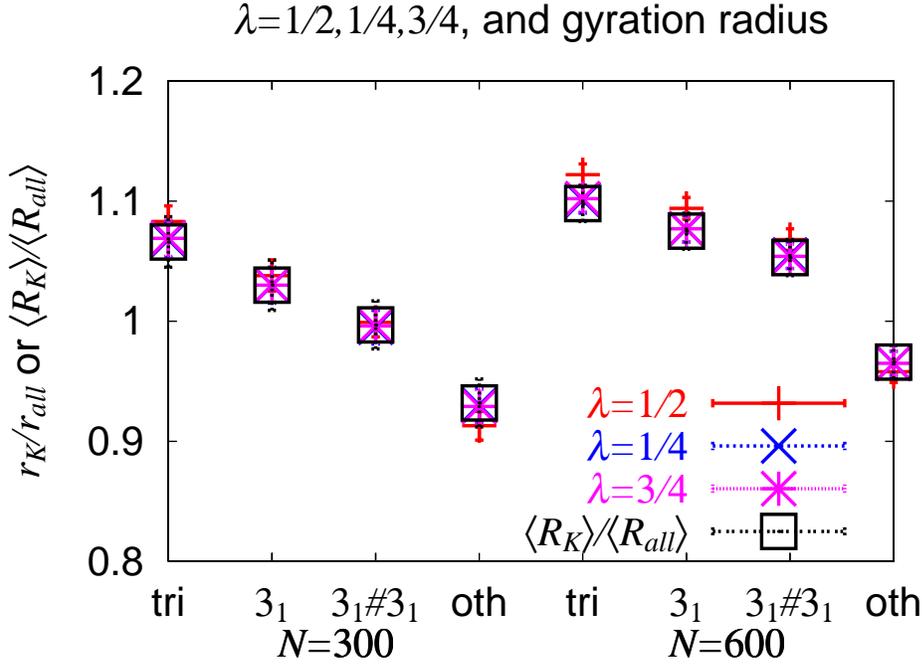}
    	\caption{The ratio of the average distances $r_K/r_{all}$ 
            and that of the gyration radii $\langle R_K \rangle/\langle R_{all} \rangle$ 
            for a given topological condition $K$.  
            The ratios $r_K(\lambda,N)/r_{all}(\lambda,N)$ for 
    	    $\lambda$ = 1/2, 1/4, and 3/4 are denoted by dots($\centerdot$), 
            crosses($\times$), and double crosses($\ast$), respectively.  
            Here  $K$ is given by 
            $\emptyset$, $3_1$, $3_1 \sharp 3_1$, and $others$({\bf oth}). 
            The ratio $\langle R_K \rangle/\langle R_{all} \rangle$  
            is denoted by open squares($\square$) for the four cases of $K$.  
            Here we consider $N$ = 300 and 600.
            Dots, crosses, double crosses and squares are colored with
            red, blue, fuchsia and black, respectively.
            }
	\label{fig:ratio}
\end{figure}

 We now show that the $\lambda$- and $N$-dependence of the average distance 
$r_K(\lambda,N)$ is given by the Gaussian one 
in the cases of $\lambda=$ 1/4, 1/2 and 3/4 for the three knots.  
Let us denote by $P({\bf r}; N)$ the end-to-end distance 
distribution of the Gaussian random walk of $N$ steps. 
For random polygons consisting of two Gaussian chains of 
$\lambda N$ steps and $(1-\lambda)N$ steps, 
the probability distribution of the vector ${\bf r}$ connecting the two end-points 
should be proportional to the product 
$P({\bf r}; \lambda N)P({\bf r}; (1-\lambda)N)$. 
Here we note that the intersegment vector ${\bf r}$ 
for the parameter $\lambda$ connects the common end-points of 
the two random walks of length $\lambda N$ and $(1-\lambda)N$.  
 For the case of no topological constraint, therefore, 
the constant $D_{all}$ is given by 
\begin{equation}
D_{all} = {\frac 3 {2 \lambda (1-\lambda) N a^2}} 
\end{equation}
Here $a$ denotes the bond length, and $a=1$ in the paper.  
We thus have    
\begin{equation} 
r_{all}(\lambda, N) = \sqrt{ {\frac {8} {3 \pi}} } \,  
\sqrt{ \lambda (1-\lambda)N }  \, a 
\end{equation}
The ratio $r_K(\lambda,N)/r_{all}(\lambda,N)$ is plotted in figure~\ref{fig:ratio}  
for $N=$ 300 and 600 with respect to the topological conditions, 
$\emptyset$, $3_1$, $3_1 \sharp 3_1$, and 
$others$. In each case of  the four topological conditions, 
the ratio takes almost the same value for 
 $\lambda=$ 1/2, 1/4 and 3/4. Furthermore, we find that the ratio should also coincide 
with the ratio of the gyration radii, $R_K/R_{all}$. 
Thus, we have the conjecture: $r_K(\lambda,N) =  r_{all}(\lambda, N) {R_K}/{R_{all}}$. 
If the conjecture is true, then the $\lambda$-dependence of 
the distribution $f_K(r, \lambda, N)$ is completely given by the Gaussian one, 
and the $N$-dependence is given by the Gaussian with the rescaling factor 
${R_K}/{R_{all}}$.  Here we remark that the ratio  ${R_K}/{R_{all}}$ may depend on 
the number of nodes $N$, since  random polygons under a topological constraint can be 
larger or smaller than that of no topological constraint 
due to entropic repulsion induced by the topological constraint.

We explain some details about the average size of random polygons. 
We first recall that the symbol $R_K$ denotes 
the square root of the mean square radius of gyration 
for random polygons of a topological constraint $K$. 
We denote by $\langle R_K \rangle$ 
the mean gyration radius of a polygon 
averaged over an ensemble of random polygons of a topological constraint $K$. 
In figure~\ref{fig:ratio}, the ratio  
$\langle R_K \rangle/\langle R_{all} \rangle$ is plotted 
for the four topological conditions.   
However, the difference between the two ratios, 
$R_K/R_{all}$ and $\langle R_K \rangle/\langle R_{all} \rangle$, 
should be smaller than the error bars.

\begin{figure}[htbp]
    	\includegraphics[width=130mm]
{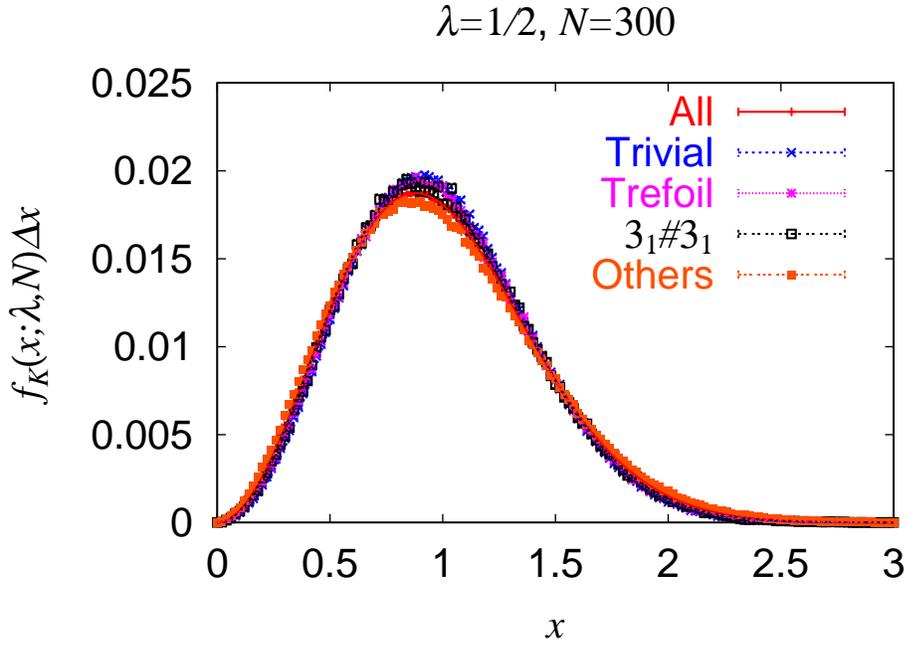}
    	\caption{Distribution ${\tilde f}_K(x; \lambda,N)$ 
    	of normalized intersegment distance $x=r/r_K$ 
    	for $\lambda = 1/2$ and $N = 300$. 
    	%Here $K$ denotes the five topological conditions.
	The distributions ${\tilde f}_K(x; \lambda,N)$ for $all$,
        $\emptyset$, $3_1$, $3_1\sharp 3_1$, and $others$ are denoted
        by dots($\centerdot$), crosses($\times$), double crosses($\ast$), 
        open squares($\square$) and closed squares($\blacksquare$), respectively.
        The plots and fitting curves are colored with 
        red, blue, fuchsia, black and orange, respectively.
 Here $\Delta r$ of eq. (\ref{eq:dist_inter_K}) is given by $\Delta x = 0.02$.  
 }
		\label{fig:knot}
\end{figure}

Let us discuss the knot-dependence of the distribution $f_K$. 
We show that the distribution ${\tilde f}_K$ of normalized distance $x$ 
for a knot $K$ should be almost independent of the knot type. 
 In figure \ref{fig:knot} the rescaled distribution ${\tilde f}_K$ 
for the five topological conditions 
are plotted against the normalized intersegment distance $x=r/r_K$ 
for the case of $N=300$ and $\lambda=1/2$.  
We see in figure~\ref{fig:knot} that   
the distributions ${\tilde f}_K$ should almost the same for the three knots. 
It is consistent with the observation that 
the estimates of the parameters are of similar values 
for the five topological conditions, as shown in tables 
 \ref{table:refit_result_1/2} and \ref{table:refit_result_1/4}.   
 Thus, the knot-dependence of the distribution $f_K$ should be 
renormalized into the value $r_K(\lambda,N)$.

%Let us  summarize the results in the subsection.  We have found that the distribution 
%$f_K(r;\lambda,N)$ is well fitted by the same functional form as the Gaussian one. 
%However, the position of the peak depends on the knot type $K$. 
%For simple knots such as the $\emptyset$, $3_1$ and $3_1 \sharp 3_1$,
%the peak positions are shifted to larger values than that of the
%distribution $f_{all}$ of polygons under no topological constraint. 

%%%%%%%%%%%%%%%%%%%%%%%%%%%%%%%%%%%%%%%%%%%%%%%%
%\newpage 

\subsection{Distribution $g_K(R;N)$ of gyration radius $R$ }

\begin{figure}[htbp]
    	\includegraphics[width=130mm]
{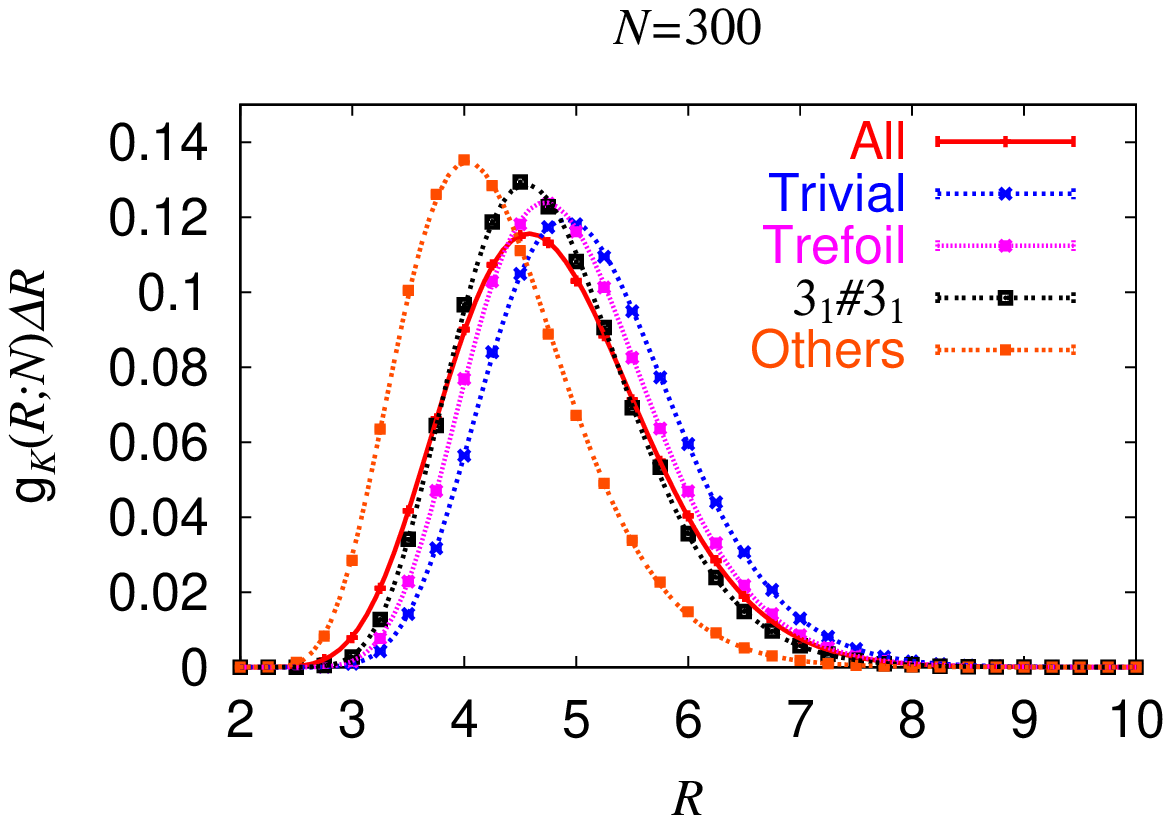}
    	\caption{Distribution $g_K(R;N)$ 
    	of gyration radius $R$ for $N = 300$.
	Dots($\centerdot$), crosses($\times$), double crosses($\ast$), 
        open squares($\square$) and closed squares($\blacksquare$)
	denote the distribution $g_K(R;N)$ for $all$,
        $\emptyset$, $3_1$, $3_1\sharp 3_1$ and $others$, respectively.
        The plots and fitting curves are colored with 
        red, blue, fuchsia, black and orange, respectively.
}
		\label{fig:dist_size_300}
\end{figure}

\begin{figure}
    	\includegraphics[width=130mm]
{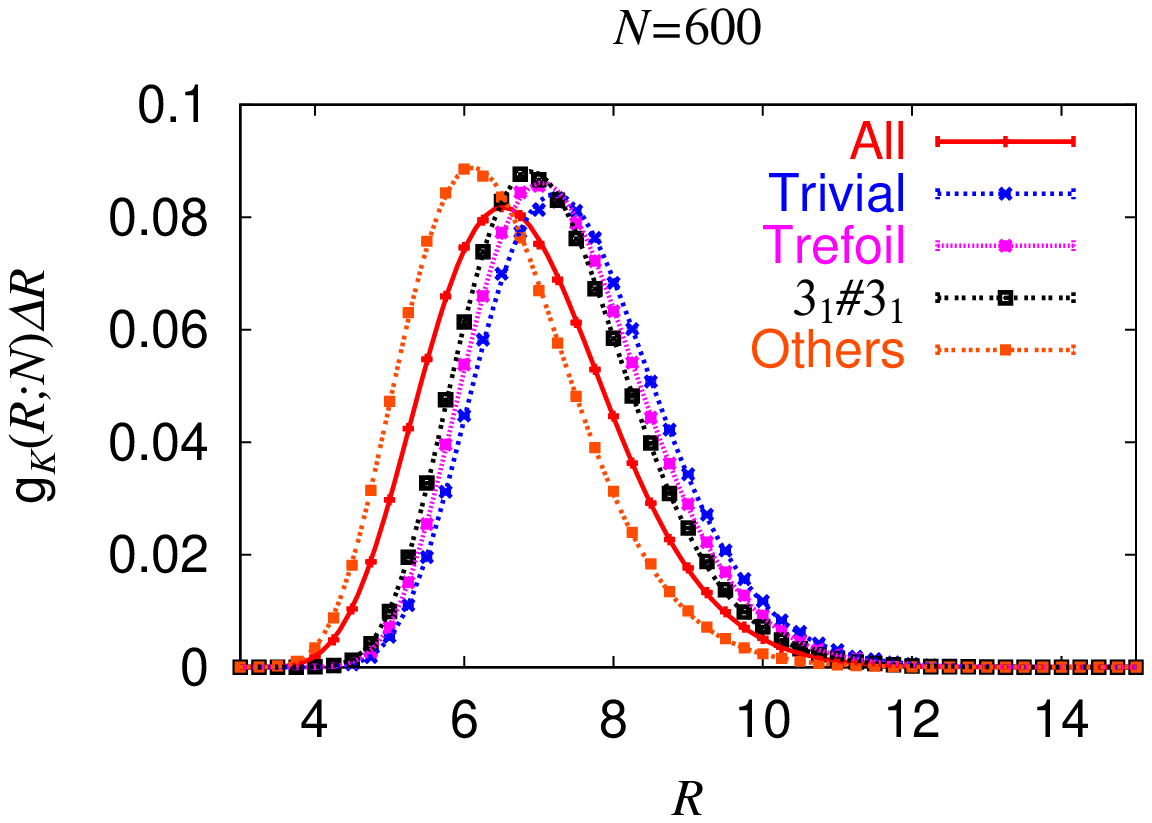}
    	\caption{Distribution $g_K(R;N)$ of gyration radius 
	$R$ for $N = 600$.
	Dots($\centerdot$), crosses($\times$), double crosses($\ast$), 
        open squares($\square$) and closed squares($\blacksquare$)
	denote the distribution $g_K(R;N)$ for 
        $all$, $\emptyset$, $3_1$, $3_1\sharp 3_1$, and $others$, respectively.
        The plots and fitting curves are colored with
        red, blue, fuchsia, black and orange, respectively.}
	\label{fig:dist_size_600}
\end{figure}

The distribution functions $g_K(R;N)$ 
of gyration radius $R$ are shown in 
figures~\ref{fig:dist_size_300} and \ref{fig:dist_size_600}  
with respect to the five topological conditions 
for $N = 300$ and $600$, respectively.     
Here the five cases are the same as  in \S 4.2. We denote 
the distributions $g_K$ for the five cases briefly   
as $g_{all}$, $g_\emptyset$, $g_{3_1}$, $g_{3_1 \sharp 3_1}$,
and $g_{others}$, respectively. 
The fitting curves in figures~\ref{fig:dist_size_300} 
and \ref{fig:dist_size_600} fit well to the data points 
in some ranges of $R$.   
The curves are given by the formula (\ref{eq:scaling_dist_size_K}).   
Thus, the formula (\ref{eq:scaling_dist_size_K}) 
approximates the distribution $g_K(R;N)$ of gyration radius 
effectively. They could be useful for studying topological effects 
on the gyration radius.

Let us consider plotting the distribution $g_K$ with respect to 
a normalized variable of the gyration radius, $R/R_K$. 
More precisely, we consider plotting $g_K$ in terms of 
the variable $R/\langle R_K \rangle$. Here we recall 
that $\langle R_K \rangle$ denotes the 
average of the gyration radius of a polygon 
averaged over an ensemble of random polygons with a given knot $K$. 
We introduce distribution 
${\tilde g}_K(y; N)$ of the variable $y=R/\langle R_K \rangle$ by the relation:    
 ${\tilde g}_K(y; N) dy = g_K(R; N) dR$. Hereafter, however, 
we denote $\langle R_K \rangle$ simply by $R_K$, except for figure captions.

\begin{figure}[htbp]
    	\includegraphics[width=130mm]
{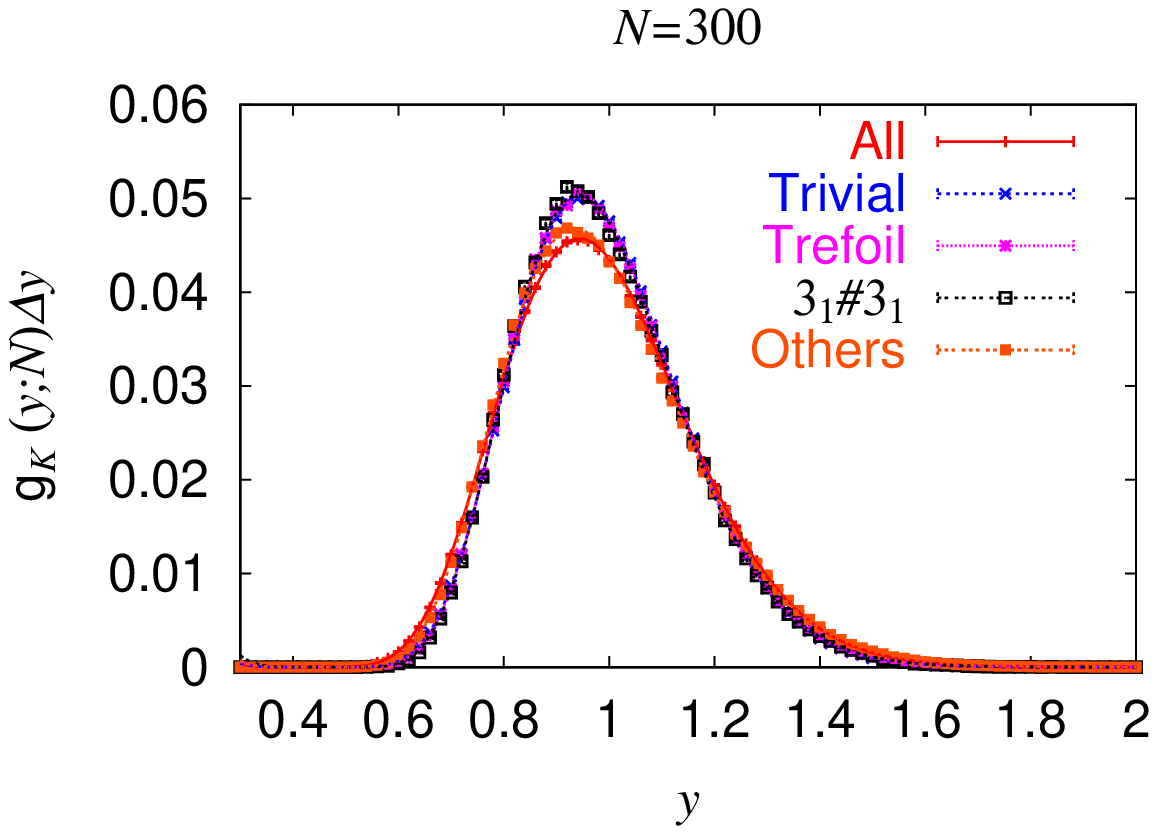}
    	\caption{Distribution ${\tilde g}_K(y; N)$ 
    	of the normalized gyration radius $y=R/\langle R_K \rangle$ for $N = 300$.
	Dots($\centerdot$), crosses($\times$), double crosses($\ast$), 
        open squares($\square$) and closed squares($\blacksquare$)
	denote the distribution ${\tilde g}_K(y; N)$ for the conditions,
        $all$, $\emptyset$, $3_1$, $3_1\sharp 3_1$, and $others$, respectively.
        The plots and fitting curves are colored with
        red, blue, fuchsia, black and orange, respectively.
 Here $\Delta R$ of eq. (\ref{eq:dist_size_K}) is given by $\Delta y = 0.02$.  
}
		\label{fig:normalized_300}
\end{figure}

\begin{figure}
    	\includegraphics[width=130mm]
{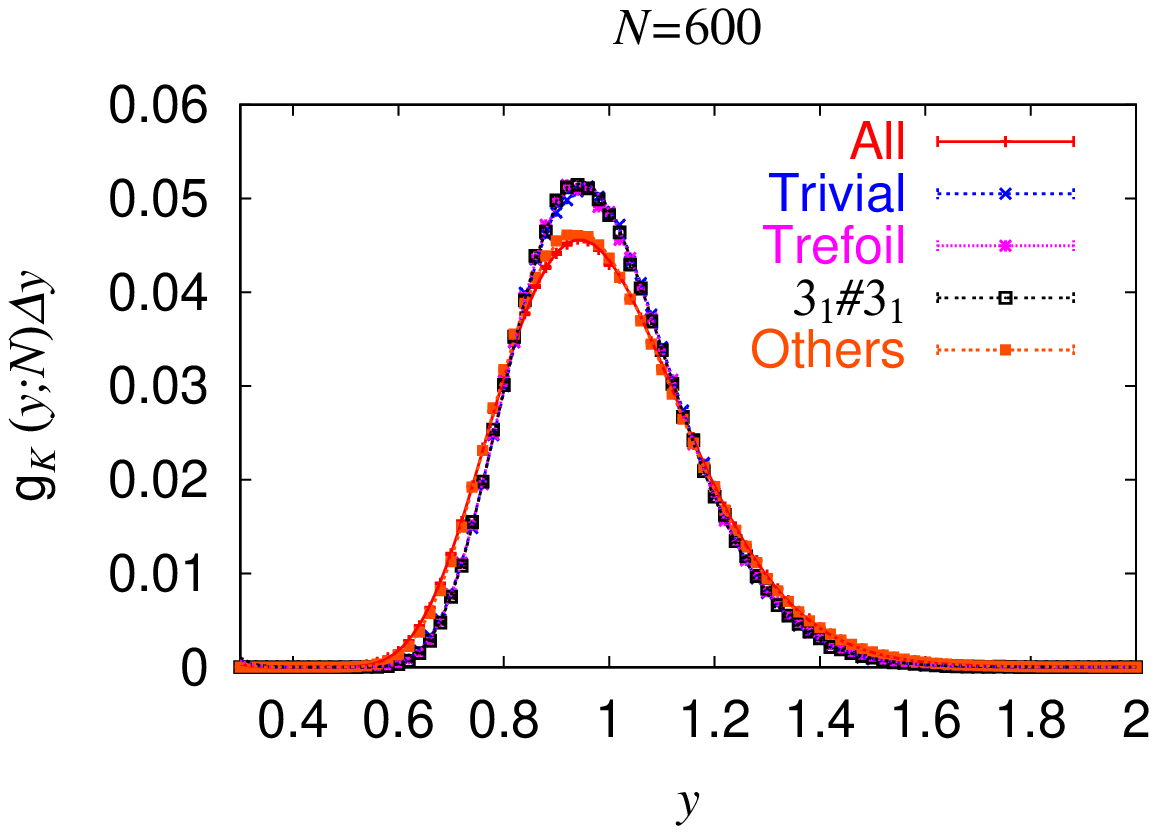}
    	\caption{Distribution ${\tilde g}_K(y;N)$ of 
    	the normalized gyration radius 
        $y=R/R_K$ for $N = 600$.
	Dots($\centerdot$), crosses($\times$), double crosses($\ast$), 
        open squares($\square$) and closed squares($\blacksquare$)
	denote the distribution ${\tilde g}_K(y;N)$ for the conditions, 
        $all$, $\emptyset$, $3_1$, $3_1\sharp 3_1$, and $others$, respectively.
        The plots and fitting curves are colored with
        red, blue, fuchsia, black and orange, respectively.
 Here $\Delta R$ of eq. (\ref{eq:dist_size_K}) is given by $\Delta y = 0.02$.  
}
		\label{fig:normalized_600}
\end{figure}

We now present the rescaled distribution ${\tilde g}_K$  
in figures~\ref{fig:normalized_300} and \ref{fig:normalized_600} 
 for $N=300$ and $N=600$, respectively. 
We find that for a given knot $K$ 
the distribution ${\tilde g}_K$ of normalized gyration radius $y=R/R_K$  
should be independent of the knot type almost completely.   
In figures~\ref{fig:normalized_300} and \ref{fig:normalized_600},   
the data points and their fitting curves of ${\tilde g}_K$ 
overlap each other for the cases of the three knots, $\emptyset$, $3_1$, 
and $3_1 \sharp 3_1$.  
We also find that the fitting curves are fit well to 
the data points both in figures~\ref{fig:normalized_300} 
and \ref{fig:normalized_600}. They are drawn by a fitting formula 
corresponding to  (\ref{eq:scaling_dist_size_K})  
\begin{equation} 
{\tilde g}_K(y; N) = {\tilde A}_{g,K} |y - {\tilde c}_K|^{\theta_{g,K}}
	\exp \left[ -{\tilde D}_{g,K} |y - {\tilde c}_K|^{\delta_{g,K}} \right] 
\label{eq:normalized_dist_size_K}
\, . 
\end{equation}
The fitting parameters 
 ${\tilde A}_{g,K}$, ${\tilde D}_{g,K}$ and ${\tilde c}_K$ correspond  
 to ${A}_{g,K}$, $D_{g,K}$ and ${c}_K$ of 
 the formula (\ref{eq:scaling_dist_size_K}) as  
\begin{equation}
A_{g,K}= {\tilde A}_{g,K} /R_K^{1 + \theta_{g,K}} \, , 
\quad 
D_{g,K} = {\tilde D}_{g,K} /R_K^{\delta_{g,K}} \, , 
\quad 
c_K = {\tilde c}_K \, R_K \, . 
 \end{equation}
The best estimates of the fitting parameters are listed 
in table~\ref{table:size_dist}. 
The $\chi^2$ values are good, in particular,  
 for the cases of the three knots, 
  ${\tilde g}_{\emptyset}$, ${\tilde g}_{3_1}$ and ${\tilde g}_{3_1 \sharp 3_1}$. 
Here the fitting range of $y=R/R_K$ is given by from 0.4 to 2.0 
for all the fitting parameters given in table~\ref{table:size_dist}.

The knot-dependence and the $N$-dependence
 of distribution $g_K$ should be renormalized 
into the mean square radius of gyration, $R^2_K(N)$. 
The rescaled distributions ${\tilde g}_K$ for 
the three knots do not depend 
on the polygonal length, $N$. The fitting curves 
of ${\tilde g}_K$ for the three knots are almost the same 
for $N=300$ and $N=600$. 
We can confirm the observation in figures~\ref{fig:normalized_300} 
and \ref{fig:normalized_600} 
by comparing the estimates shown in  table~\ref{table:size_dist}.  
The fitting parameters ${\theta}_{g,K}$, 
${\delta}_{g,K}$, ${\tilde A}_{g,K}$ 
and ${\tilde c}_{K}$ depend on neither 
the knot type $K$ nor the polygonal length $N$ 
with respect to their errors.    
The normalization of gyration radius, $R/R_K$,  
should be thus essential when analyzing 
 the distribution function $g_K$.

 Random polygons of relatively simple knots should be larger in size 
than the average one, while those of more complex knots should be smaller.    
It depends on the polygonal length $N$ whether 
the size of random polygons of a given knot should be 
larger or smaller than the average.   
In figure~\ref{fig:dist_size_300}, 
the peak position of the distribution $g_{\emptyset}$ is larger than those of the  
other distributions $g_{all}$, $g_{3_1}$, $g_{3_1 \sharp 3_1}$ 
and $g_{others}$. In figure~\ref{fig:dist_size_600}, 
the peaks of $g_{\emptyset}$, $g_{3_1}$, $g_{3_1 \sharp 3_1}$
are clearly located on the right hand side of the peak of $g_{all}$, while 
the peak of $g_{others}$ is located on the left hand side. 
The equilibrium length of a random knot 
can be a criterion whether it is larger or smaller than the average  
~\cite{DIAO-AKOS-KUSNER-MILLETT-STASIAK.2003.1}.  
We also note that the peak positions of distributions $g_K$ shown in 
figures~\ref{fig:dist_size_300} and \ref{fig:dist_size_600}  
are roughly consistent with the average values of $R_K$~\cite{DEUTSCH.1999.01,SHIMAMURA-DEGUCHI.2001.03,MATSUDA-YAO-TSUKAHARA-DEGUCHI-FURUTA-INAMI.2003.01,DOBAY-DUBOCHET-MILLET-SOTTAS-STASIAK.2003.01}.

\section{Discussion}

%Conclusion and Discusion
%Concl.tex
%2004/02/16 Created
%2004/02/19 Last Modified Date

We have found that the distribution $f_K$ of intersegment distance for 
random polygons under topological constraint $K$ is almost given by the Gaussian 
distribution. 
Furthermore, rescaling the distance by the average distance $r_K$, 
we have shown that the $\lambda$- and $N$-dependence of $f_K$ is renormalized 
into the average distance $r_K(\lambda, N)$.  
We have proposed the conjecture: $r_K=r_{all} \, R_K/R_{all}$, 
for any $\lambda$, $N$ and $K$. Here we have assumed that 
$N$ is large enough. 
If it is true, then topological constraints do  
not have any effect on the distribution 
of intersegment distance, $f_K$, except for scaling 
the distance by the factor $R_K/R_{all}$.

The effect of the ``topological excluded volume'' 
should  be rather different from the standard excluded volume effect 
of self-avoiding walks. It does not correspond to a real excluded volume,   
although the ratio $R_K/R_{all}$ of a knot $K$ 
can become larger than 1 in the case of large $N$.   
When $N$ increases, random polygons with more complex knots can appear,   
which should be smaller than those of a simple knot. 
If we consider only such random polygons that have a fixed simple knot, 
then the size can be larger than the average one when $N$ is very large. 
Topological constraints thus may induce effective swelling of random polygons. 
However, they do not change the functional form of the 
distribution of the distance between two segments.

Let us discuss the difference in terms of critical exponents. 
We denote by $\nu_K'$ 
the scaling exponent defined for the asymptotic
behavior of the average size of SAW such as given in
(\ref{eq:size_scaling}).  
For SAW, the exponent $\nu_K'$ corresponds to 
the exponent $\nu_K$ determined by the formula (\ref{eq:scaling_large_dist_inter_K}) 
~\cite{DES_CLOIZEAUX.1974.01}.
If des Cloizeaux's relations (\ref{eq:theta_large_dist_inter_K}) and (\ref{eq:delta_large_dist_inter_K}) 
could be valid for  random polygons under topological constraints, 
we should have $\nu_K' \simeq 0.50$ from the best estimates for the distribution $f_K(r; \lambda, N)$
as shown in table~\ref{table:fit_result_dist_inter}.

 Within the scope of the present research, however, 
it is not clear whether the two exponents $\nu_K$ and $\nu_K'$ should be equal or not.  
Moreover, it is not clear whether the relations (\ref{eq:theta_large_dist_inter_K}) 
and (\ref{eq:delta_large_dist_inter_K}) 
should be valid for random polygons under topological constraints.  
It seems that the form of the distribution $f_K(r;\lambda,N)$ remains Gaussian with
the exponent $\nu_K \simeq 0.50$ in the limit $N \rightarrow \infty$. 
However,  the average size $R^2_K(N)$ might follow the scaling form 
with a different exponent, $\nu_K' > 0.5$.

\section*{Acknowledgment}
The authors are grateful to M. K. Shimamura for helpful discussions.
A.Y. was supported in part by the Chuo University.	

\bibliographystyle{unsrt}
%\bibliography{PAP.bbl}
%\bibliography{polymer}

%%%%%%%%%%%%%%%%%%%%
% Fig.           %
%%%%%%%%%%%%%%%%%%%%
%Figures
%Fig.tex
%2004/06/06 Created
%2004/06/06 Last Modified Date

%%%%%%%%%%%%%%%  FIG 1 %%%%%%%%%%%%%%%%%%%

%%%%%%%%%%%%%%%  FIG 2 %%%%%%%%%%%%%%%%%%%

%%%%%%%%%%%%%%%  FIG 3 %%%%%%%%%%%%%%%%%%%

%%%%%%%%%%%%%%%  FIG 4 %%%%%%%%%%%%%%%%%%%

%%%%%%%%%%%%%%%  FIG 5 %%%%%%%%%%%%%%%%%%%

%%%%%%%%%%%%%%%%%%%%%%% Fig 6 %%%%%%%%%%%%%%%%%%%%%%%%%%%%%

%%%%%%%%%%%%%%%%%%%%%%% FIG 7 %%%%%%%%%%%%%%%%%%%%%%%%%%%%%%%%%%%%%

%%%%%%%%%%%%%%%%%%%%%%% FIG 8 %%%%%%%%%%%%%%%%%%%%%%%%%%%%%%%%%%%%%

%%%%%%%%%%%%%%%%%%%%%%% FIG 9 %%%%%%%%%%%%%%%%%%%%%%%%%%%%%%%%%%%%%

%%%%%%%%%%%%%%%%%%%%%%%%%%%%%%

%%%%%%%%%%%%%%%%%%%%
% Table.           %
%%%%%%%%%%%%%%%%%%%%

%%%%%%%%%%%%%%%%%%%%% TAB 1 %%%%%%%%%%%%%%%%%%%%%%
\begin{table}[htpb]
{\small
\begin{center}
\caption{The fitting values of the scaling formula
 (\ref{eq:scaling_large_dist_inter_K}) 
 to the data of the distribution $f_K(r;\lambda,N)$ at $\lambda = 1/2$
 with $\chi^2$ values per datum.
%$N$ and $K$ denote the polygonal length and 
%the topological conditions, respectively. 
%The $\chi^2$ value per datum is shown in the column of $\chi^2$. 
}
\label{table:fit_result_dist_inter}
$\begin{array}{cccccccc}
\hline
\hline
N & K & \gamma_{K} & \nu_{K} & D_{K} \times 10^{2} 
	& A_{K} \times 10^{3} 
	& \chi^2 & \shortstack[10mm]{Fitting\\ Range}\\
\hline
300 & all & 1.04 \pm 0.03 & 0.503 \pm 0.005 
	& 1.9 \pm 0.1 & 1.78 \pm 0.09 & 1.13 & 5.75-20\\
    & \emptyset & 0.87 \pm 0.13 & 0.509 \pm 0.019 
	& 1.7 \pm 0.5 & 0.8 \pm 0.1 & 1.05 & 6.5-16\\
    & 3_1 & 0.93 \pm 0.09 & 0.50 \pm 0.01 
	& 1.9 \pm 0.3 & 1.1 \pm 0.1 & 0.90 & 6.25-19\\
    & 3_1\#3_1 & 1.0 \pm 0.1 & 0.51 \pm 0.02 
	& 2.0 \pm 0.6 & 1.5 \pm 0.3 & 0.85 & 5.75-17\\
    & others & 1.1 \pm 0.1 & 0.48 \pm 0.02
	& 2.9 \pm 0.6 & 4.0 \pm 0.5 & 0.81 & 5.5-15\\
\hline
\hline
600 & all & 1.03 \pm 0.02 & 0.501 \pm 0.004 
	& 0.99 \pm 0.05 & 0.63 \pm 0.03 & 0.76 & 7.5-30\\
    & \emptyset & 0.8 \pm 0.2 & 0.51 \pm 0.03 
	& 0.8 \pm 0.4 & 0.19 \pm 0.07 & 0.78 & 9-23\\
    & 3_1 & 0.8 \pm 0.2 & 0.51 \pm 0.02 
	& 0.9 \pm 0.3 & 0.22 \pm 0.06 & 1.22 & 8.5-23.75\\
    & 3_1\#3_1 & 1.0 \pm 0.1 & 0.52 \pm 0.02 
	& 0.7 \pm 0.3 & 0.29 \pm 0.06 & 0.92 & 7.5-22\\
    & others & 1.06 \pm 0.05 & 0.493 \pm 0.008 
	& 1.2 \pm 0.1 & 0.94 \pm 0.09 & 0.97 & 8.25-28.5\\
\hline
\end{array}$
\end{center}}
\end{table}

%%%%%%%%%%%%%%%%%%   TAB 2   %%%%%%%%%%%%%%%%%%%%%%%%%%
\begin{table}[htpb]
{\small
\begin{center}
\caption{The fitting values of the scaling formula
 (\ref{eq:scaling_large_dist_inter_K}) to the data 
 of the re-scaled distribution ${\tilde f}_K(x; \lambda,N)$ at $\lambda = 1/2$
 with $\chi^2$ values per datum.
%$N$ and $K$ denote the polygonal length and 
%the topological condition, respectively. 
%The $\chi^2$ value per datum is shown in the column of $\chi^2$. 
}
\label{table:refit_result_1/2}
$\begin{array}{cccccccc}
\hline
\hline
N & K & \gamma_{K} & \nu_{K} & D_{K} 
	& A_{K} \times 10^{2} 
	& \chi^2 & \shortstack[10mm]{Fitting\\ Range}\\
\hline
300 & all & 1.04 \pm 0.04 & 0.504 \pm 0.006 
	& 1.25 \pm 0.04 & 6.3 \pm 0.2 & 0.80 & 0.70 - 2.00\\
    & \emptyset & 0.89 \pm 0.08 & 0.51 \pm 0.01 
	& 1.39 \pm 0.08 & 7.7 \pm 0.6 & 1.20 & 0.70 - 2.00\\
    & 3_1 & 0.99 \pm 0.08 & 0.52 \pm 0.01 
	& 1.30 \pm 0.08 & 6.9 \pm 0.5 & 0.89 & 0.70 - 2.00\\
    & 3_1\#3_1 & 1.1 \pm 0.1 & 0.52 \pm 0.02 
	& 1.2 \pm 0.1 & 6.4 \pm 0.8 & 0.90 & 0.70 - 2.00\\
    & others & 1.04 \pm 0.07 & 0.49 \pm 0.01
	& 1.26 \pm 0.07 & 6.2 \pm 0.4 & 0.91 & 0.70 - 2.00\\
\hline
\hline
600 & all & 1.03 \pm 0.04 & 0.502 \pm 0.006 
	& 1.26 \pm 0.04 & 6.4 \pm 0.2 & 0.75 & 0.70 - 2.00\\
    & \emptyset & 1.0 \pm 0.1 & 0.53 \pm 0.02 
	& 1.3 \pm 0.1 & 7.2 \pm 0.8 & 0.81 & 0.70 - 2.00\\
    & 3_1 & 0.9 \pm 0.1 & 0.51 \pm 0.02 
	& 1.4 \pm 0.1 & 7.8 \pm 1.0 & 1.26 & 0.70 - 2.00\\
    & 3_1\#3_1 & 1.0 \pm 0.1 & 0.52 \pm 0.02 
	& 1.3 \pm 0.1 & 7.2 \pm 1.0 & 1.05 & 0.70 - 2.00\\
    & others & 1.06 \pm 0.05 & 0.499 \pm 0.008 
	& 1.24 \pm 0.05 & 6.2 \pm 0.3 & 1.00 & 0.70 - 2.00\\
\hline
\end{array}$
\end{center}}
\end{table}

%%%%%%%%%%%%%%%%%%% TAB 3 %%%%%%%%%%%%%%%%%%%%%
\begin{table}[htpb]
{\small
\begin{center}
\caption{The fitting values of the scaling formula
 (\ref{eq:scaling_large_dist_inter_K}) 
 to the data of the re-scaled distribution 
 ${\tilde f}_K(x; \lambda,N)$ at $\lambda = 1/4$
 with $\chi^2$ values per datum.
%$N$ and $K$ denote the polygonal length and 
%the topological condition, respectively. 
%The $\chi^2$ value per datum is shown in the column of $\chi^2$. 
}
\label{table:refit_result_1/4}
$\begin{array}{cccccccc}
\hline
\hline
N & K & \gamma_{K} & \nu_{K} & D_{K} 
	& A_{K} \times 10^{2} 
	& \chi^2 & \shortstack[10mm]{Fitting\\ Range}\\
\hline
300 & all & 1.10 \pm 0.04 & 0.514 \pm 0.006 
	& 1.19 \pm 0.04 & 6.0 \pm 0.2 & 0.88 & 0.70 - 2.00\\
    & \emptyset & 1.01 \pm 0.09 & 0.52 \pm 0.01 
	& 1.28 \pm 0.08 & 6.8 \pm 0.6 & 1.54 & 0.70 - 2.00\\
    & 3_1 & 1.15 \pm 0.07 & 0.54 \pm 0.01 
	& 1.14 \pm 0.07 & 5.9 \pm 0.4 & 0.94 & 0.70 - 2.00\\
    & 3_1\#3_1 & 1.0 \pm 0.2 & 0.51 \pm 0.02 
	& 1.3 \pm 0.2 & 6.8 \pm 1.0 & 1.06 & 0.70 - 2.00\\
    & others & 1.11 \pm 0.07 & 0.50 \pm 0.01
	& 1.20 \pm 0.07 & 5.8 \pm 0.4 & 1.03 & 0.70 - 2.00\\
\hline
\hline
600 & all & 1.05 \pm 0.05 & 0.507 \pm 0.007 
	& 1.24 \pm 0.05 & 6.2 \pm 0.3 & 1.28 & 0.70 - 2.00\\
    & \emptyset & 1.0 \pm 0.1 & 0.53 \pm 0.02 
	& 1.3 \pm 0.1 & 7.0 \pm 0.7 & 0.78 & 0.70 - 2.00\\
    & 3_1 & 0.8 \pm 0.1 & 0.50 \pm 0.02 
	& 1.4 \pm 0.1 & 8.1 \pm 1.1 & 1.24 & 0.70 - 2.00\\
    & 3_1\#3_1 & 0.9 \pm 0.2 & 0.51 \pm 0.02 
	& 1.4 \pm 0.1 & 7.5 \pm 1.1 & 1.18 & 0.70 - 2.00\\
    & others & 1.13 \pm 0.05 & 0.511 \pm 0.008 
	& 1.17 \pm 0.05 & 5.8 \pm 0.3 & 1.07 & 0.70 - 2.00\\
\hline
\end{array}$
\end{center}}
\end{table}

%%%%%%%%%%%%%%%%%%%% TAB 4 %%%%%%%%
%	Gyration Radius Table

\begin{table}
{\small
\begin{center}
\caption{The fitting values of the formula (\ref{eq:normalized_dist_size_K})
to the data of the re-scaled distribution ${\tilde g}_K(y; N)$ of 
the normalized gyration radius $y=R/\langle R_K \rangle$ 
for $N$-noded random polygons of topological condition $K$
with $\chi^2$ values per datum. 
The fitting range of $y$ is from 0.4 to 2.0. 
%The $\chi^2$ value per datum is shown in the column of $\chi^2$. 
}
\label{table:size_dist}
$\begin{array}{cccccccc}
\hline
\hline
N & K & \theta_{g,\,K} & \delta_{g,\,K} 
	& {\tilde D}_{g,\,K} & {\tilde A}_{g,\,K} \times 10^{-4}& {\tilde c}_K
	& \chi^2 \\
\hline
300 & all & 6.2 \pm 0.3 & 1.44 \pm 0.03 & 11.0 \pm 0.4
	& 0.019 \pm 0.009 & 0.423 \pm 0.007 & 3.63 \\
    & \emptyset & 7.9 \pm 0.4 & 1.31 \pm 0.03 & 14.6 \pm 0.6 
	& 0.4 \pm 0.3 & 0.441 \pm 0.006 & 1.61 \\
    & 3_1 & 8.3 \pm 0.3 & 1.21 \pm 0.03 & 16.1 \pm 0.6 
	& 1.7 \pm 1.1 & 0.450 \pm 0.004 & 3.33 \\
    & 3_1\#3_1 & 8.7 \pm 0.5 & 1.13 \pm 0.03 & 17.3 \pm 0.8 
	& 6 \pm 5 & 0.457 \pm 0.005 & 1.16 \\
    & others & 7.7 \pm 0.3 & 1.12 \pm 0.03 & 15.1 \pm 0.6 
	& 1.1 \pm 0.6 & 0.441 \pm 0.004 & 3.97 \\
\hline
\hline
600 & all & 6.2 \pm 0.3 & 1.42 \pm 0.02 & 11.0 \pm 0.3 
	& 0.020 \pm 0.007 & 0.424 \pm 0.006 & 5.91 \\
    & \emptyset & 7.7 \pm 0.3 & 1.30 \pm 0.03 & 14.9 \pm 0.4 
	& 0.46 \pm 0.23 & 0.457 \pm 0.005 & 1.86 \\
    & 3_1 & 8.3 \pm 0.3 & 1.21 \pm 0.03 & 16.3 \pm 0.5 
	& 1.9 \pm 1.1 & 0.458 \pm 0.004 & 2.39 \\
    & 3_1\#3_1 & 8.1 \pm 0.3 & 1.22 \pm 0.03 & 16.0 \pm 0.5 
	& 1.3 \pm 0.6 & 0.463 \pm 0.003 & 2.38 \\
    & others & 7.0 \pm 0.3 & 1.25 \pm 0.03 & 13.1 \pm 0.5
	& 0.14 \pm 0.08 & 0.434 \pm 0.006 & 6.90 \\
\hline
\end{array}$
\end{center}}
\end{table}

%%%%%%%%%%%%%%%%%%%%%%%%%%%%
%	Normalized Factor Table
%\begin{table}[htpb]{\begin{center}
%\caption{The average of the intersegment distance, $r_K(\lambda, N)$, 
%for $\lambda=1/2$ and 1/4, and the average of the gyration radius 
%of random polygons uner topological constraint $K$, ${\bar R}_K$. }
%\label{table:factor}
%$\begin{array}{ccccc}\hline \hline
%N & K & r_K \, (\lambda = 1/2) & r_K \, (\lambda = 1/4) & {\bar R}_K  \\ \hline
%300 & all & 7.997 \pm 0.005 &6.892 \pm 0.004 & 4.932 \pm 0.003\\
%    & \emptyset & 8.661 \pm 0.009 & 7.358 \pm 0.008 & 5.258 \pm 0.006\\
%    & 3_1 & 8.302 \pm 0.010 & 7.102 \pm 0.009 & 5.078 \pm 0.006\\
%    & 3_1\#3_1 & 7.988 \pm 0.017 & 6.860 \pm 0.014 & 4.916 \pm 0.010\\
%    & others & 7.301 \pm 0.007 & 6.411 \pm 0.006 & 4.594 \pm 0.004\\ \hline\hline
%600 & all & 11.293 \pm 0.007 & 9.759 \pm 0.006 & 6.967 \pm 0.004\\
%    & \emptyset & 12.669 \pm 0.024 & 10.743 \pm 0.021 & 7.652 \pm 0.015\\
%    & 3_1 & 12.355 \pm 0.020 & 10.512 \pm 0.017 & 7.488 \pm 0.012\\
%    & 3_1\#3_1 & 12.065 \pm 0.023 & 10.299 \pm 0.020 & 7.338 \pm 0.014\\
%    & others & 10.814 \pm 0.008 & 9.420 \pm 0.007 & 6.732 \pm 0.005\\
%\hline \end{array}$ \end{center}} \end{table}

\end{document}